# Giant Topological Hall Effect in van der Waals Heterostructures of CrTe$_2$/Bi$_2$Te$_3$

Xiaoqian Zhang[†‡%#], Siddhesh C. Ambhire[¶#], Qiangsheng Lu[‡#], Wei Niu[§], Jacob Cook[‡], Jidong Samuel Jiang[∥], Deshun Hong[∥], Laith Alahmed[⊥], Liang He[†], Rong Zhang[†], Yongbing Xu*[†▽], Steven S.-L. Zhang*[¶], Peng Li*[⊥], Guang Bian*[‡]

[†]Jiangsu Provincial Key Laboratory of Advanced Photonic and Electronic Materials, School of Electronic Science and Engineering, Nanjing University, Nanjing 210093, China

[‡]Department of Physics and Astronomy, University of Missouri, Columbia, Missouri 65211, USA

[%]Shenzhen Institute for Quantum Science and Engineering, Southern University of Science and Technology, Shenzhen, Guangdong, 518055, China

[¶]Department of Physics, Case Western Reserve University, Cleveland, Ohio 44106, USA,

[§]New Energy Technology Engineering Laboratory of Jiangsu Provence & School of Science, Nanjing University of Posts and Telecommunications, Nanjing 210023, China

[∥]Materials Science Division, Argonne National Laboratory, Lemont, IL 60439, USA

[⊥]Department of Electrical and Computer Engineering, Auburn University, Auburn, AL 36849, USA

[▽]York-Nanjing Joint Centre (YNJC) for spintronics and nano engineering, Department of Electronic Engineering, The University of York, YO10 3DD, United Kingdom

Abstract:

Discoveries of interfacial topological Hall effect (THE) provide an ideal platform for exploring physics arising from the interplay between topology and magnetism. The interfacial topological Hall effect is closely related to the Dzyaloshinskii-Moriya interaction (DMI) at interface and topological spin textures. However, it is difficult to achieve a sizable THE in heterostructures due to the stringent constraints on the constituents of THE heterostructures such as strong spin-orbit coupling (SOC). Here we report the observation of a giant THE signal of 1.39 μΩ·cm in the van der Waals heterostructures of $CrTe_2/Bi_2Te_3$ fabricated by molecular beam epitaxy, a prototype of two-dimensional (2D) ferromagnet (FM)/topological insulator (TI). This large magnitude of THE is attributed to an optimized combination of 2D ferromagnetism in $CrTe_2$, strong SOC in $Bi_2Te_3$, and an atomically sharp interface. Our work reveals $CrTe_2/Bi_2Te_3$ as a convenient platform for achieving large interfacial THE in hybrid systems, which could be utilized to develop quantum science and high-density information storage.

Over the past decade, real-space magnetic chiral spin textures have generated enormous attention for featuring Berry curvature physics and inherent magnetic stability.[1-3] Generally, the topological spin texture with scalar spin chirality in real-space provides an emergent field, which acts on charge carriers and consequently generates the topological Hall effect (THE).[4, 5] This real-space THE is distinct from the intrinsic anomalous Hall effect (AHE) in ferromagnets (FMs), which results from the Berry phase in momentum space. Magnetic skyrmions were firstly observed in noncentrosymmetric chiral-lattice magnets like B20-type MnSi/MnGe[6-8] and $Fe_{0.5}Co_{0.5}Si$,[9] which showed prominent THE responses in transport measurements. Since then, the THE has been considered as an experimental signature of the topological spin texture of skyrmions.

The occurrence of the delicate spin texture of skyrmions depends on the strength of the Dzyaloshinskii-Moriya interaction (DMI). A practical way to generate DMI is to construct heterostructures with spin ordering and strong spin-orbit coupling (SOC). Inversion symmetry is always broken at the interface of two dissimilar materials in a heterostructure, which is essential for generating DMI. Advances in fabricating artificial structures based on thin-film deposition techniques enable construction of heterostructures with essential ingredients for interfacial DMI and skyrmions. Néel-type skyrmions have been probed in hybrid systems, including FM/heavy-metal (HM),[2, 10-12] FM/topological insulator (TI),[13-15] TI/magnetic insulator (MI)[16] and HM/MI heterostructures.[5, 17] These hybrid systems with magnetic skyrmions are summarized in Fig. 1. For instance, a two-dimensional (2D) skyrmion lattice was revealed in monolayer Fe on Ir surface, originating from the co-interplay of four-spin interaction and DMI.[12] Moreover, interfacial DMI and skyrmions have been observed in $(BiSb)_2Te_3$/GdFeCo heterostructures,[15] benefiting from the strong SOC in TIs. Among all the research framework of interfacial hybrid systems with interfacial magnetic skyrmions, 2D FM/TI, a van der Waals (vdW) heterostructure, remains largely unexplored. Compared with conventional hybrid systems involving three-dimensional (3D) metals, vdW heterostructures hold many advantages, including reduced sample dimension (with thickness down to a few atomic layers), flexibility in stacking order, and the atomically sharp interface between vdW components. Those features favor the formation of strong interfacial DMI and chiral spin textures such as skyrmions.

Therefore, optimization of vdW heterostructures provides a promising opportunity for exploring interfacial skyrmion physics.

2D vdW magnets with intrinsic ferromagnetic order preserved in monolayer limit provide opportunities for probing the THE. The strong uniaxial magnetic anisotropy ($K$) of 2D magnets is vital for the formation of 2D ferromagnetism. However, it also raises the threshold of DMI for generating the chiral spin texture ($D_c = 2\sqrt{2JK}/\pi$, where $D_c$ represents the critical DMI value and $J$ is the exchange interaction coefficient).[18] TI can serve as the other component in a bilayer heterostructure to provide strong SOC that can modulate spins in real space.[15] In addition, the topological surface states (SS) of TI could mediate the interfacial DMI.[19, 20] It has been demonstrated that Néel-type skyrmions exist in mechanically exfoliated 1T'-$WTe_2$/$Fe_3GeTe_2$ heterostructures.[21] Very recently, room-temperature intrinsic ferromagnetism has been observed in $CrTe_2$ epitaxial films,[22] indicating that $CrTe_2$ is a promising spin host for THE heterostructures. Furthermore, the epitaxial thin films are critically significant for spintronics device applications as they offer the opportunity to fabricate wafer-size heterostructures and superlattices with thickness down to atomic layers. As a prototypical TI, $Bi_2Te_3$ features a single Dirac cone, which can act as nontrivial SS to mediate DMI.[23-25] Moreover, compared with Ta, W and Pt compounds, $Bi_2Te_3$ possesses a much larger SOC,[26, 27] due to the facts that Bi is the heaviest nonradioactive element and the atomic SOC is proportional to the fourth power of atomic number, $Z^4$. Besides robust spin order in $CrTe_2$ and strong SOC in $Bi_2Te_3$, $CrTe_2$ and $Bi_2Te_3$ also share three crucial features for realizing strong interfacial DMI: 1) vdW interlayer coupling, 2) simple telluride compounds, and 3) hexagonal surface lattice symmetry, which facilitates the formation of an atomically sharp interface between the two compounds. Therefore, $CrTe_2$/$Bi_2Te_3$ heterostructure presents an optimal condition for stabilizing interfacial THE.

Here, we report the realization of an optimal THE heterostructure, $CrTe_2$/$Bi_2Te_3$, as demonstrated by angle-resolved photoemission spectroscopy (ARPES) and scanning tunneling microscopy (STM) and magneto-transport measurements. The epitaxial heterostructure was grown by molecular beam epitaxy (MBE) with thickness down to a few atomic layers. A pronounced THE resistivity of ~ 1.39 μΩ·cm was observed at 10 K, much larger than the typical THE strength ~0.1 μΩ·cm of previous THE bilayer heterostructures. The observed THE signals in $CrTe_2$/$Bi_2Te_3$ persist up to 100 K, indicating a highly robust topologically non-trivial chiral spin texture at the interface.

## Results and Discussion

**2D FM/TI heterostructures.** 10 Quintuple layers (QLs) of $Bi_2Te_3$ thin films were firstly grown on sapphire ($Al_2O_3$) substrates (see Methods), followed by 10 trilayers (TLs) of $CrTe_2$ on top. 1T-$CrTe_2$ has a layered trigonal crystal structure with the space group of $p\bar{3}m1$ and lattice constants $a$=3.8 Å and $c$=6.1 Å.[28] A schematic of $CrTe_2$/$Bi_2Te_3$ bilayer is shown in Figure 2a, in which the red arrows represent the magnetic moments of Cr. Few-layer $CrTe_2$ is a ferromagnetic material with strong perpendicular magnetic anisotropy (PMA). The magnetic moments of Cr atoms in adjacent layers are ferromagnetically coupled with an atomic value of ~3 $\mu_B$/atom. 2D crystalline thin films were characterized by STM (Figure 2c-2f). $Bi_2Te_3$ and $CrTe_2$ surfaces are identified *via* comparisons of their respective step edge heights and atomic lattice distances. Figure 2c and 2e are the large-scale STM images of atomically flat $Bi_2Te_3$ and $CrTe_2$ thin films with height profiles (inset) showing different layers. The hexagonal atomic lattices of the top Te layer of $Bi_2Te_3$ and $CrTe_2$ with a spacing of 0.44 nm and 0.38 nm are shown in Fig. 2d and 2f, respectively. A typical layer-by-layer growth mode of $CrTe_2$/$Bi_2Te_3$ bilayers with atomically flat surfaces and the same Te terminated layers endows the system with emergent phenomenon and tunable functionality.

ARPES measurements were performed to visualize the band structure of the heterostructures. The band spectra of $Bi_2Te_3$ and $CrTe_2$ thin films along M–Γ–M direction are shown in Figure 2g and 2h, respectively. The "M" shape bulk valence band (BVB) of $Bi_2Te_3$ locates at the bottom. The clear linear band dispersion confirms the massless Dirac SS, which do not disperse with photon energy (see Figure S2). The Dirac point is located at ~0.25 eV below the Fermi level. These typical features are in agreement with previous works.[23, 29] In the spectrum of $CrTe_2$, two hole-pockets locate near the Fermi level. These two pockets are from the majority and minority spin bands of $CrTe_2$, which is consistent with the first-principles results.[22] The spin band structure confirms the intrinsic ferromagnetism in $CrTe_2$ thin layers. The coexistence of intrinsic topological SS with strong SOC and magnetic states at the interface establishes that the $CrTe_2$/$Bi_2Te_3$ bilayer is a viable platform to study the THE.

**Berry phase effect in real space: THE.** The $CrTe_2$/$Bi_2Te_3$ bilayers were patterned into a Hall bar geometry for magnetotransport measurements. The schematic diagram of the experimental setup is illustrated in Figure 3a. Figure 3b shows the Hall resistivity ($\rho_{xy}$) as a function of the magnetic field. At low temperatures, the hysteresis loops show a pronounced square shape, which is a hallmark of robust ferromagnetism with strong out-of-plane magnetic anisotropy. The remanent $\rho_{xy}$ and coercive field ($H_C$) vanish at around 280 K, indicating a Curie temperature ($T_C$) near this temperature.

The positive slope at high field suggests dominant hole carriers, echoing the electronic structure of $CrTe_2$ where the hole pockets in the valence band cross the Fermi level. Apart from that, an anomaly in $\rho_{xy}$ is noticed below 120 K. There is a sharp antisymmetric peak near $H_C$ followed by a smooth suppression at the high field region. The emergent hump and dip features indicate an additional contribution to the Hall signal, which shares similarity with the THE response. When the spin-polarized electrons at the interface pass through a chiral spin texture, they pick up the real-space Berry phase by aligning their spins with the local magnetic moments, giving rise to the THE contribution to the Hall resistivity.

In general, the Hall resistivity of a ferromagnetic conductor consists of three contributions: an ordinary Hall effect (OHE) term due to the Lorentz force acting on the charge carriers, the AHE term proportional to the magnetization and the THE term resulting from the real-space Berry phase. Therefore, $\rho_{xy}$ can be expressed as:

$$\rho_{xy}(H) = R_0 H + R_s M + \rho_{THE},$$

where $R_0$ is the ordinary Hall coefficient determined from a linear fitting of $\rho_{xy}$ at high magnetic fields, $R_s$ is the anomalous Hall coefficient of the ferromagnetic component, and $\rho_{THE}$ is the topological Hall contribution. The THE and AHE contributions should be extracted to understand the temperature evolution of THE and AHE. Two main approaches are employed to single out THE component. The first one is using a step function, $M_0 \tanh\left(\frac{H}{a_0} - H_0\right)$ to extract the AHE contribution.[18] At a high magnetic field, all the spins align ferromagnetically, leading to the absence of spin chirality. Under this circumstance, $\rho_{xy}$ is only from the AHE and ordinary Hall term. When the magnetization starts to change orientation near $H_C$, the $\rho_{xy}$ curve deviates from AHE behavior and develops a broad hump over AHE loops. An example of the AHE background subtraction based on the fitting function of $M_0 \tanh\left(\frac{H}{a_0} - H_0\right)$ is exhibited in Figure 3c. The topological Hall term is highlighted by green areas. The other method is taking the difference of resistivity between upward and downward field scans.[30]

Considering the positive field region at 50 K, the $CrTe_2/Bi_2Te_3$ system is in a topologically trivial ferromagnetic state without any spin chirality during the downward field sweep. When the sweep direction changes from downward to upward, the system evolves into an intermediate chiral spin texture during the process of magnetic transitions. Therefore, $\rho_{THE}$ can be extracted by taking the difference between the hysteresis loops under upward and downward field sweeps. We have used both methods to extract the THE contribution, and the fitting results are equivalent. The temperature-dependent contributions from AHE and THE are shown in Figure S3.

**Temperature and field evolution of AHE and THE.** After subtracting the OHE and THE term, the net AHE signal is displayed in Figure 3d. Nonlinear field dependence of $\rho_{AHE}$ with a negative value emerges at 280 K. Upon cooling, a pronounced hysteresis loop appears, indicating the formation of ferromagnetic order. In particular, the AHE changes its polarity as temperature decreases below 50 K (Figure 3e), at which $\rho_{xy}$ is dominated by the THE contribution. We now discuss possible origins of the polarity change of AHE in $CrTe_2/Bi_2Te_3$. Since the magnetization (M) is positive at a high positive magnetic field, the negative AHE is definitely due to the sign-change of $R_s$. $R_s$ is the ratio of the anomalous Hall resistivity to the M, which consists of contributions from extrinsic dynamical processes such as skew scattering and side-jump, and intrinsic Berry curvature of the band structure. The extrinsic mechanisms depend on the impurities, scattering time, and density of states at the Fermi level.[31] On the other hand, the intrinsic AHE is related to the integral of the Berry curvature over the Fermi sea.[32] Similar sign-change behavior of AHE was previously observed in ferromagnetic $SrRuO_3$ crystals and $Cr_x(Bi_{1-y}Sb_y)_{2-x}Te_3/(Bi_{1-y}Sb_y)_2Te_3$ heterostructures, which is theoretically shown to result from the magnetic monopoles in momentum space[32] and Rashba splitting of bulk bands,[33] respectively. In addition, temperature-dependent sign-change of $\rho_{AHE}$ has been reported in Co/Pd multilayers,[34] in which the phonons/magnons scattering has a crucial contribution. The variation of momentum-space Berry curvature around the Fermi level is also a possible explanation for the polarity change of AHE in $CrTe_2/Bi_2Te_3$ at low temperatures. Considering that the longitudinal resistivity of $CrTe_2$ increases with temperature, the enhanced $\rho_{AHE}$ with temperature is likely due to the growing electron-phonon or magnons scattering. When the temperature further increases above 120 K, $\rho_{AHE}$ is reduced due to the suppressed magnetization at high temperatures. The interplay of ferromagnetism, topology and spin-dependent scattering leads to complex AHE behaviors.

The magnetic-field dependence of the THE shows a hysteresis behavior in Figure 4a. At 10 K, THE reaches its maximum at 7.9 kOe and decays gradually to zero at high fields. The THE signal persists up to 100 K, despite the decreased intensity at high temperature. Above this temperature, only AHE is present. Note that the THE features with a similar critical temperature can be reproduced in another $CrTe_2/Bi_2Te_3$ bilayer sample (see Figure S4). To further study the characteristics of the Hall anomaly, we extracted the peak value of $\rho_{THE}$ ($\rho_{THE}^{max}$), $H_C$ and peak position of humps ($H_T$) in Figure 4b. $\rho_{THE}^{max}$, the maximal amplitude of THE signal, manifests a pronounced value of 1.39 μΩ·cm at 10 K, and decreases monotonically at elevated temperatures. Notably, the observed $\rho_{THE}^{max}$ in $CrTe_2/Bi_2Te_3$ is the largest among interfacial and bulk skyrmion systems known to date, including Ir/Fe/Co/Pt (0.03 μΩ·cm),[11] $SrIrO_3/SrRuO_3$ (0.2 μΩ·cm),[35] $Pt/Y_3Fe_5O_{12}$ (0.3 μΩ·cm)[36] and bulk MnSi (0.04 μΩ·cm),[7] as summarized in Table 1. We note that the THE amplitude observed in this work is lower than the magnetic bubbles system $Ca_{0.99}Ce_{0.01}MnO_3$ (120 μΩ·cm).[4] The amplitude of the topological Hall resistivity reflects the strength of the coupling between electric current and the spin structure.[37] For metallic systems, the reduced electron mean free path and the large charge carrier density usually lead to a significant reduction of THE contribution to Hall resistivity because the conductivity is sensitive to the disorder.[38] However, in $CrTe_2/Bi_2Te_3$ systems, the nontrivial SS of TI provides nearly dissipationless spin transport, which enhances the electron mean free path. Additionally, the atomically sharp interface between vdW layers is also favorable for enhancing interfacial DMI and hence the THE.

The temperature dependence of $H_T$ and $H_C$ are summarized in Figure 4c. The curve of $H_T$ at which $\rho_{THE}$ reaches its maximum follows the trend of $H_C$, suggesting that the spin chirality is induced when the magnetic moments of Cr atoms start to be reversed. Note that the THE response is developed when carriers pass through the emergent magnetic field, that is, a fictitious field derived from the Berry phase in real-space. The topological Hall feature is the strongest at low temperature, then monotonically decreases at elevated temperatures and disappears above 100 K. Although this critical temperature is lower than that of magnetic skyrmion multilayers composed of heavy-metal thin films (usually above room temperature), it is significantly higher than typical

TI-based skyrmion systems, like MnTe/(Bi,Sb)$_2$Te$_3$ (20 K),[14] Mn-Bi$_2$Te$_3$ (14 K)[19] and Cr$_2$Te$_3$/Bi$_2$Te$_3$ (20 K).[13] It is comparable with other vdW interfacial DMI systems, such as Fe$_3$GeTe$_2$/WTe$_2$ (100 K).[21] Moreover, the sign of THE remains positive in the entire temperature window, irrespective of the sign change of AHE. It indicates that the origin of THE is entirely different from that of AHE. The THE is associated with the DMI strength, while AHE is related to the resistivity and magnetization.

How electric currents can manipulate chiral spin textures in hybrid systems remains elusive. To investigate it, experimental depinning and motion of spin chirality by applying current pulses were conducted in the CrTe$_2$/Bi$_2$Te$_3$ system, as exhibited in Figure 4e. Upon passing a current of $-j_e$ through the device, the left of the device develops dense chiral spin textures, giving rise to an increased Hall voltage, and vice versa. This is a strong evidence of skyrmions as similar Hall voltage change was observed when current was used to pulse skyrmions.[38] Since the current can only shrink and elongate the topologically trivial magnetic textures, it clearly follows that the topologically nontrivial spin textures in CrTe$_2$/Bi$_2$Te$_3$ system with well-defined chirality can be efficiently manipulated, including depinning and motion. In order to rule out the Joule heating effect, we have carried out the control experiment with pulse width of 0.01 ms (see Figure S5). The curve is almost the same as the case for the 1 ms pulse width. These results show that Joule heating has a negligible effect on the current-induced skyrmion depinning and motion experiment.[39]

**Quantitative agreement between transport and theoretical simulations.** Combining the THE signals for all temperatures and magnetic fields, a phase diagram is plotted in Figure 5a. As temperature decreases, the THE amplitude increases and can exist in a stronger magnetic field. It is because THE is induced by the thermodynamic stability of the chiral spin texture, which is enhanced by lowering the temperature. Therefore, a stronger magnetic field is needed to eliminate this nontrivial spin texture. The large amplitude of THE at low temperatures manifests a stronger fictitious magnetic field induced by Berry phase effects in real space. The emergence of THE humps across a wide range of temperature from 10 K to 100 K, suggests that the topological spin structure is robust. The THE effect exists only below the $T_C$ of CrTe$_2$, which rules out the possibility of spin chirality driven by thermal fluctuation.[18]

Moreover, the THE is absent in the transport properties of pure $CrTe_2$ thin films,[22] implying the important role of TI in the formation of spin chirality.

To understand spin chirality and the experimental phase diagram in $CrTe_2/Bi_2Te_3$ hybrid system, we calculated the energy landscapes of various magnetic states, which arise from the competition between exchange coupling, uniaxial anisotropy, and DMI in this system.[40] Figure 4c shows a magnetic phase diagram in the plane of magnetic field and temperature calculated by using zero-temperature saturation magnetization ($M_0 = 362$ emu/cm$^3$), experimental perpendicular anisotropy coefficient ($K_u = 3.16 \times 10^5$ J/m$^3$), and taking the DMI coefficient[41] as $D(T) = D_0 \left[1 - \left(\frac{T}{T_c}\right)^{\frac{3}{2}}\right]^5$ with $D_0 = 6 \times 10^{-3}$ J/m$^{-2}$ and the $T_C$ of $CrTe_2$ being 280 K. The Ginzburg-Landau model[42] is employed in the calculation (Supporting Information Section 7). At low magnetic fields, a helical phase with zero net out-of-plane magnetization is energetically favorable. When the out-of-plane magnetic field is increased to a window of intermediate field strength, a skyrmion lattice phase prevails over the helical phase as the out-of-plane magnetization in the skyrmion core areas lowers the Zeeman energy. Further increasing the magnetic field leads to a ferromagnetic phase (Figure 5d) with magnetization uniformly aligned with the external magnetic field.

To compare with the experimental phase diagram for the THE resistivity, we also calculated topological charge density as a function of the magnetic field and temperature (Figure 5b) based on the same materials parameters used in the calculation of Figure 5c. Comparing with Figure 5a, we find the calculated magnetic-field window for finite topological charge density is relatively narrower, which could be attributed to the following reasons. In our theoretical calculation, we only considered closely-packed hexagonal skyrmion lattice whereas in reality there may exist other topological spin structures such as a disordered skyrmion phase[43] that may contribute to the measured THE as well. In addition, derivation from the linear dependence of the topological Hall resistivity on the topological charge density may take place in the presence of strong SOC[44] and diffusive scattering.[45] Except for the width of THE window on magnetic field axis for a particular value of temperature, the calculated diagram of topological charge density is in good agreement with the experimental phase diagram of THE.

The spin-chiral structure responsible for the observed interfacial THE is most likely the Néel-textured magnetic skyrmions stabilized by DMI. In contrast to Bloch-type skyrmions in bulk DMI materials with $\rho_{THE}$ typically between 5~100 nΩ·cm and narrow temperature-field range near $T_C$,[6, 7] Néel-type skyrmions usually exhibit a more pronounced $\rho_{THE}$ and a wider temperature-field range,[46] as seen in $CrTe_2/Bi_2Te_3$. Similar THE behaviors induced by Néel-type magnetic skyrmions were also observed in $Cr_2Te_3/Bi_2Te_3$,[13] $Pt/TmIG$,[5] $La_{0.7}Sr_{0.3}MnO_3/SrIrO_3$[47] and $(Bi,Sb)_2Te_3/MnTe$ heterostructures.[14]

In order to rule out the possibility of two-channel AHE responses,[48, 49] which could lead to similar hump-dip features in Hall resistivities, we did two-channel AHE fitting using a double-tanh function (see Figure S6). A simple linear superposition of two magnetic hysteresis loops can't match the $\rho_{xy}$-$H$ curves, especially for the hump-like features. It suggests that the hump features are not due to the mixed magnetic signals from two AHE responses with different $H_C$.

**Calculation of skyrmion size.** In skyrmion systems, reducing the skyrmion size is critical to maximizing its potential for device applications of energy-efficient and high-density information storage. A single skyrmion can be treated as one magnetic flux quantum, $\phi_0 = h/e$, where $h$ is the Plank constant and $e$ is the electronic charge. When charge carriers flow through a conductor, the skyrmion density ($n_{sk}$) gives rise to an emergent electromagnetic field. Therefore, $\rho_{THE}$ can be represented by the following formula[38]

$$\rho_{THE} = PR_0 n_{sk} \phi_0,$$

where $R_0$ is the ordinary Hall coefficient representing the total density of mobile charge and $n_{sk}$ is the 2D skyrmions density (assuming each skyrmion carries a topological charge $|Q_{sk}| = 1$, and skyrmions form a regular 2D lattice). $P$ denotes the spin polarization of carriers. Assuming $P = 1$, we can estimate the length scale of one single skyrmion ($n_{sk}^{-1/2}$) to be around 34 nm, which is the upper limit of skyrmion size. This value is comparable to that of bulk B20 alloys[6-8, 50] and $SrRuO_3$-$SrIrO_3$ bilayers.[35, 51] Since the skyrmion density is proportional to the magnitude of the emergent magnetic field, skyrmions with smaller size are essential to optimize the relationship between spin dynamics and electric charge. Most recently, magnetic bubbles have aroused researchers' enthusiasm owing to the observation of an extremely large THE in $(Ca,Ce)MnO_3$,[4] ten thousand times larger than the THE in MnSi.[7] However, the size of magnetic bubbles is typically around 0.1-10 $\mu$m, which is too large for practical applications in information storage. Unlike magnetic bubbles, the skyrmions stabilized by the DMI scale around sub-100-nm, either in bulk or interface. Here, the estimated maximum value of skyrmion size in $CrTe_2/Bi_2Te_3$ system is ~34 nm, much smaller than the size of magnetic bubbles. Furthermore, skyrmions can easily be moved by a low-threshold electric current/field, while magnetic bubbles require a large external magnetic field to manipulate.

The observed THE magnitude is the largest among all the bilayer systems hosting skyrmions.[6, 11, 35, 36, 52] The pure vdW staking in $CrTe_2/Bi_2Te_3$ heterostructures explored in this work is totally different from the traditional magnetic bilayer systems. The vdW coupling between adjacent layers in $CrTe_2/Bi_2Te_3$ heterostructures facilitates the layer-by-layer stacking and the formation of a clean interface, without suffering from interfacial hybridization and diffusion, strain, surface reconstruction and electronic redistribution.[53] Common Te atoms in both of TI and 2D FM permit the topological SS to extend into the 2D FM. Combined with the out-of-plane spin-polarized $CrTe_2$ surface, efficient magnetic exchange coupling could be introduced at the $CrTe_2/Bi_2Te_3$ interface, giving rise to a gapped TI SS and the dissipationless chiral edge conductive channel. As the DMI is largely determined by magnetic interactions within a single atomic layer, the atomically sharp vdW interface together with anisotropic exchange interaction and dissipationless spin carrier, as well as the strong SOC in $Bi_2Te_3$ provide an optimal environment for the formation of interfacial DMI. The DMI energy parameter can be defined as: $E_{DMI} = Dt$ with $D = 6\times10^{-3}$ J/m$^{-2}$ and $t$ (6 nm) representing the thickness of $CrTe_2$ layer. This leads to an enhanced DMI strength in $CrTe_2/Bi_2Te_3$ of 36 pJ/m, which is one or two orders larger than that of traditional heavy-metal skyrmion systems such as Pt/TmIG (0.02 pJ/m)[5] and Ir/Co/Pt (0.96 pJ/m).[54] These results indicate that the interfacial DMI at 2D FM/TI heterostructures is one of the promising ways to produce large THE, and therefore a higher recording density.

## Conclusions

In summary, we have synthesized high-quality $CrTe_2/Bi_2Te_3$ bilayers, a prototype vdW 2D FM/TI heterostructure by MBE, which shows the largest THE among bilayer systems. It is driven by an enhanced interfacial DMI as a consequence of an optimal combination of strong SOC in $Bi_2Te_3$, and an atomically sharp interface with asymmetric exchange interaction. It is also worth noting that the 2D FM/TI hybrid system possesses topologically nontrivial spin textures in both real space and momentum space, namely, chiral spin textures and spin-polarized topological SS of TI. This work provides a platform to explore and manipulate the interplay among magnetism, spin textures, and Berry phase in a class of unexplored bilayers for the development of energy-efficient nanoscale spintronic devices.

**Experimental Methods**

The deposition of $CrTe_2/Bi_2Te_3$ heterostructures on $Al_2O_3$(0001) substrates were carried out in a MBE chamber with a base pressure of $2\times10^{-10}$ mbar. Prior to sample growth, substrates were raised to 800 °C, and left for 30 mins to outgas. $Bi_2Te_3$ thin films were firstly grown on $Al_2O_3$ at a low temperature of ~240 °C. During the growth of $CrTe_2$ thin films, the $Bi_2Te_3/Al_2O_3$ substrates were maintained at 350 °C. We note that the surface morphology of $Bi_2Te_3$ films after annealing at 350 °C shows domains with a step height of ~0.4 nm (see the Supplementary Information).[13] It suggests the existence of Bi bilayers on the surface of $Bi_2Te_3$ films, which is beneficial to the enhanced SOC at the $CrTe_2/Bi_2Te_3$ interface.

To avoid possible contamination, a 10 nm thick Te layer was deposited on $CrTe_2/Bi_2Te_3/Al_2O_3$ structures at room temperature prior to taking samples out of the MBE chamber for *ex-situ* measurements.

The ARPES spectra were collected with a SPECS PHOIBOS 150 hemisphere analyzer with a SPECS UVS 300 helium discharge lamp at 107 K. The He I$\alpha$ (21.2 eV) and He II$\alpha$ (40.8 eV) resonant lines were used to excite photoelectrons. The energy and angular resolutions were 40 meV and 0.2° under 107 K, respectively. The topography of the materials' surface was *in-situ* mapped by an Aarhus STM placed in the growth chamber.

The samples were patterned into Hall bar structures with a width of 10 $\mu$m using standard photo-lithography and an ion milling system. Ti (5 nm)/Au (150 nm) was then deposited in the Hall bar contact areas for electrical contact. The magnetic and electro-transport properties of the film were measured in a Quantum Design physical property measurement system supplemented with two lock-in-amplifiers, a preamplifier and two Keithley meters.

We have performed theoretical calculations for the free energies in the skyrmion density map based on a Ginzburg-Landau model. The skyrmion density map of magnetic field and temperature was obtained by free energy calculations using the following parameters: $D_0$ = 0.006 $J/m^3$, $T_C$ = 280 K, $M_0$ = 362 emu/cm$^3$, $k_d$ = 2.62×10$^5$ $J/m^3$, $k_u$ = 3.16×10$^5$ $J/m^3$, $A_{ex}$ = 6.5×10$^{-11}$ $J/m$, and $n$ = 2, where $D_0$, $M_0$, $k_d$ and $k_u$, represent the DMI coefficient at 0 K, saturated magnetization, demagnetizing coefficient and uniaxial anisotropy coefficient, respectively. This phase diagram was collected for various profiles of the magnetization field for the skyrmion phase which are overlaid on top of each other.

## Associated Content

**Supporting Information**

The Supporting Information is available free of charge.

Details of structural characterizations using STM and ARPES measurements, numerical fitting of Hall resistance data, extraction of the THE signals, and the calculation of the magnetic phase diagram (PDF).

## Author Information

**Corresponding Author**

*Email: ybxu@nju.edu.cn

*Email: sxz675@case.edu

*Email: pzl0047@auburn.edu

*Email: biang@missouri.edu

**Author Contributions**

[#]X.Z., S.C.A and Q.L. contributed equally.

**Notes**

The authors declare no competing financial interest.

# Figures

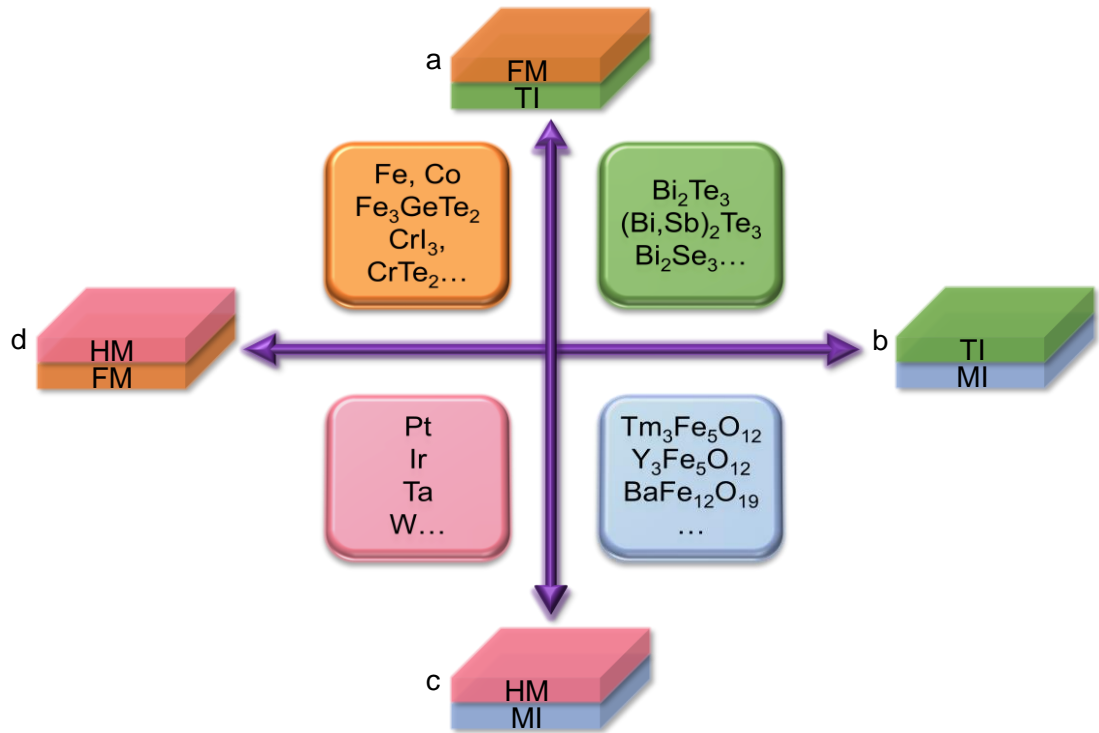

Figure 1. Research frameworks of interfacial magnetic skyrmions in hybrid systems. (a-d) Four categories of heterostructures with interfacial magnetic skyrmions, including FM/TI ($Cr_2Te_3/Bi_2Te_3$[13]) (a), TI/MI ($Bi_2Se_3/BaFe_{12}O_{19}$[16]) (b), HM/MI ($Pt/Tm_3Fe_5O_{12}$[5, 17]) (c), and HM/FM (Ir/Fe/Co/Pt[10, 11]) (d). The orange, green, blue and pink squares represent FM, TI, MI and HM, respectively.

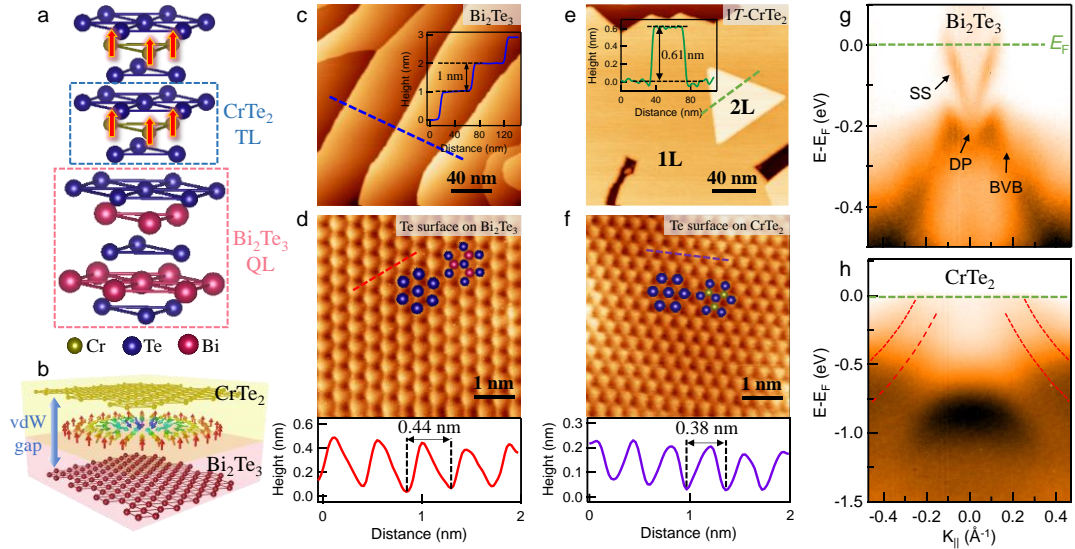

Figure 2. Atomic-scale growth and ARPES characterization of $CrTe_2/Bi_2Te_3$ heterostructure. (a) Crystal structure of the $CrTe_2/Bi_2Te_3$ heterostructure. (b) Schematic picture of $CrTe_2/Bi_2Te_3$ bilayer with vdW gap. (c,e) Surface morphologies of 10 QL $Bi_2Te_3$ (c) and 2 TL $CrTe_2$ grown on $Bi_2Te_3$ (e) (tunneling condition V = 1000 mV, I = 0.01 nA). Inset: Cross-sectional profile of QL $Bi_2Te_3$ and TL $CrTe_2$ flakes along the blue and green dashed line, respectively. (d,f) Atomically resolved Te surfaces of $Bi_2Te_3$ (d) and $CrTe_2$ (f) with corresponding atomic lattice structures (V = 10 mV, I = 0.2 nA). The panel below indicates the lattice unit distance calculated from line-cut profile. (g,h) ARPES spectra of $Bi_2Te_3$(111) (g) as well as $CrTe_2$ thin film on $Bi_2Te_3$ (h) taken with a photon energy of 21.2 eV.

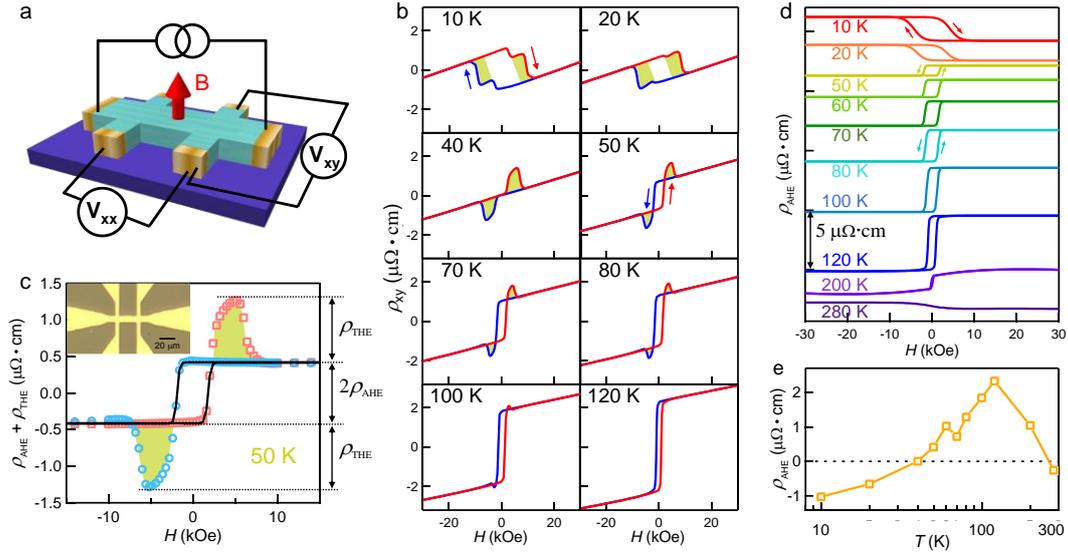

Figure 3. Hall resistivity of CrTe$_2$/Bi$_2$Te$_3$ bilayer with a perpendicular magnetic field. (a) Schematic diagram of the experimental set up for transport measurements. (b) Magnetic-field dependence of $\rho_{xy}$ at various temperatures for the CrTe$_2$/Bi$_2$Te$_3$ sample. Curves are antisymmetrized to remove the $\rho_{xx}$ component. The positive slope at high field indicates hole carriers. Red (blue) curve represents the process of increasing (decreasing) magnetic field. The humps on the shoulder of AHE loops are highlighted in light green. (c) Detailed view of the Hall resistivity at 50 K with subtraction of ordinary Hall term. Contribution from AHE (black solid lines) and THE term (light green area) is marked, where the AHE contribution is fitted by a $M_0 \tanh(\frac{H}{a_0} - H_0)$ function to represent the magnetization, where $M_0$, $a_0$ and $H_0$ is a fitting parameter. Inset: An optical image of a Hall bar device with a 20 μm scale bar. (d) Field-dependent anomalous Hall loops of the bilayer at different temperatures obtained by fitting with function of $M_0 \tanh(\frac{H}{a_0} - H_0)$. An unexpected transition of AHE component occurs with temperature decreasing below 50 K. (e) The magnitude of the AHE ($\rho_{AHE}$) at different temperatures extracted from (d).

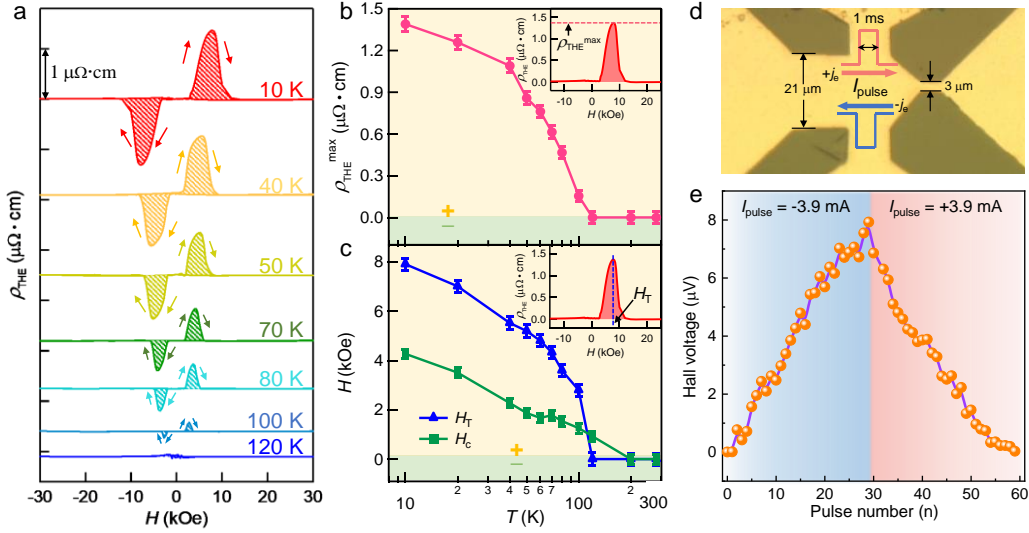

Figure 4. Interfacial coupling induced topological Hall effect and effective manipulation of magnetic skyrmions. (a) Typical pairs of magnetic-field dependence of the topological Hall resistivity at various temperatures generated after subtracting the AHE term (loops at different temperatures are offset vertically for clarity). (b,c) Temperature-dependent topological Hall effect. Red and blue solid symbols represent the amplitude ($\rho_{THE}^{max}$) and the field ($H_T$) at which the topological Hall resistivity reaches its maximum (top inset), respectively. Green squares are the $H_C$ of AHE, which scale quite well with $H_T$. (d) An optical image of a Hall bar device structure, in which a geometrical constriction is introduced for giving rise to an inhomogeneous current distribution. (e) Hall voltages recorded after repeated 1-ms-long current pulses ($j_e=\pm 3.9$ mA).

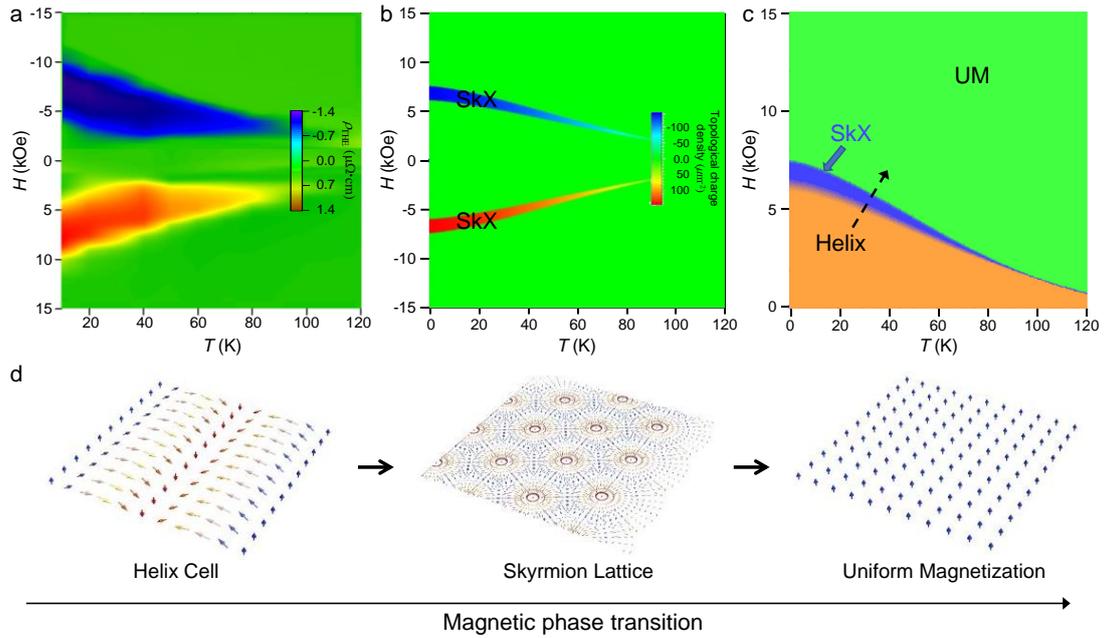

Figure 5. Topological Hall resistivity, topological charge density, magnetic phase diagram and spin ordering. (a) Experimental color map of $\rho_{THE}$ in the temperature-magnetic field plane. A larger THE amplitude at lower temperature indicates a higher density of Néel skyrmions. (b) Calculated topological charge density as a comparison with the measured topological Hall resistivity. (c) Calculated magnetic phase diagram in the plane of temperature and out-of-plane magnetic field. The colored regions correspond to the uniform FM (green), the helix (orange), and the skyrmion lattice (SkX; blue) phases, respectively. (d) Schematic spin structures of helix phases, closely-packed hexagonal SkX, and uniform ferromagnetic phase.

Table 1. Maximal THE amplitude ($\rho_{THE}^{max}$) in various skyrmion systems from literatures[α]

| Material | Maximal THE (μΩ·cm) | Ref. | System |
| --- | --- | --- | --- |
| $CrTe_2/Bi_2Te_3$ | 1.39 | [*] | Interfacial skyrmion |
| $Fe_3GeTe_2/WTe_2$ | 1.3 | 21 | Interfacial skyrmion |
| $Cr_2Te_3/Bi_2Te_3$ | 0.53 | 13 | Interfacial skyrmion |
| $Pt/Y_3Fe_5O_{12}$ | 0.3 | 36 | Interfacial skyrmion |
| $SrIrO_3/SrRuO_3$ | 0.2 | 35 | Interfacial skyrmion |
| Ir/Fe/Co/Pt | 0.03 | 11 | Interfacial skyrmion |
| $Fe_{0.7}Co_{0.3}Si$ | 0.5 | 9 | Bulk skyrmion |
| MnGe | 0.16 | 6 | Bulk skyrmion |
| MnSi | 0.04 | 7 | Bulk skyrmion |

[α] Typical values of $\rho_{THE}^{max}$ from reports. Our work [*] has demonstrated a relatively large $\rho_{THE}^{max}$ among them.